\documentclass{article}
\usepackage{amsmath,amsfonts,amssymb}
\usepackage{bbm}
\usepackage{txfonts}
\usepackage[T1]{fontenc}
\usepackage{hyperref}

\def\PL#1{{\em Phys.\ Lett.}\ {\bf#1}}

\def\PR#1{{\em Phys.\ Rev.}\ {\bf#1}}

\def\JHEP#1{{\em JHEP}\ {\bf#1}}

\def\te{\text{e}}

\def\nn{\nonumber}
\def\bt{\triangleright}

\newcommand{\arxiv}[1]{\href{https://arxiv.org/abs/#1}{arXiv:#1}}
\newcommand{\bibx}[3]{#1: #2, \arxiv{#3}}

\newcommand{\bibp}[3]{#1: #2, #3}

\begin{document}

\begin{titlepage}
\title{Generalized Heisenberg algebra, realizations of the $\mathfrak{gl}(n)$ algebra and applications}

\date{}
\clearpage\maketitle
\thispagestyle{empty}

\begin{center}

Stjepan Meljanac\\
Division of Theoretical Physics, Ru\dj er Bo\v skovi\'c Institute, Bijeni\v cka cesta 54, 10002 Zagreb, Croatia\\
meljanac@irb.hr
\bigskip
\\Zoran \v{S}koda\\
Department of Teachers' Education, University of Zadar, Franje Tu\dj mana 24, 23000 Zadar, Croatia\\
zskoda@unizd.hr
\bigskip
\\Rina \v Strajn\\
Department of Electrical Engineering and Computing, University of Dubrovnik, \'{C}ira Cari\'{c}a 4, 20000 Dubrovnik, Croatia\\
rina.strajn@unidu.hr
\end{center}

\bigskip

\begin{abstract}
We introduce the generalized Heisenberg algebra appropriate for realizations of the $\mathfrak{gl}(n)$ algebra. Linear realizations of the $\mathfrak{gl}(n)$ algebra are presented and the corresponding star product, coproduct of momenta and twist are constructed. The dual realization and dual $\mathfrak{gl}(n)$ algebra are considered. Finally, we present a general realization of the $\mathfrak{gl}(n)$ algebra, the corresponding coproduct of momenta and two classes of twists. These results can be applied to physical theories on noncommutative spaces of the $\mathfrak{gl}(n)$ type.\\

\end{abstract}

\end{titlepage}

\section{Introduction}

Noncommutative (NC) geometry appeared in theoretical physics as one possible approach towards the description of spacetime at the Planck scale \cite{dfr1,dfr2,Snyder,Luk1,Luk2}. The key assumption is noncommutativity between spacetime coordinates. The first model  was proposed by Snyder in \cite{Snyder}. There are various models of NC spaces of Lie algebra type. Specially, an interesting model is the $\kappa$-Minkowski spacetime \cite{Luk1,Luk2,majidruegg,gacluknow,dasklukwor}, where the parameter $\kappa$ is usually interpreted as the quantum gravity scale and the spacetime coordinates close a Lie algebra. The $\kappa$-Poincar\'e quantum group \cite{Luk1,Luk2,majid,ChPr}, as a possible quantum symmetry of the $\kappa$-Minkowski spacetime, allows for the study of deformed relativistic spacetime symmetries. Realizations of Lie algebras are important in the formulation of physical theories on NC spaces and the corresponding deformed symmetries. Particularly, realizations of NC coordinates are based on formal power series in the Heisenberg-Weyl algebra allowing one to simplify the methods of calculation on the deformed spacetime. Some physical applications of  Lie algebra type NC spaces and the associated star products are presented in \cite{MMMs2,GGHMM,toward}.

For certain Lie algebras, such as the orthogonal algebra $\mathfrak{so}(n)$, the Lorentz algebra $\mathfrak{so}(1,n-1)$ and the $\mathfrak{gl}(n)$ algebra,  the  realizations are not well adapted due to the structure of their commutation relations. This was the motivation for introducing the generalized Heisenberg algebra and constructing an analogue of the Weyl realization of $\mathfrak{so}(n)$ and $\mathfrak{so}(1,n-1)$ by formal power series in a semicompletion of the Heisenberg algebra \cite{MmbKJ}. This construction was applied to the extended Snyder model \cite{meljmig1} and to the unification of the $\kappa$-Minkowski and extended Snyder spaces \cite{meljmig2,meljmig3}.

In this paper we introduce another generalization of the Heisenberg algebra appropriate for realizations of the $\mathfrak{gl}(n)$ algebra. Specially, we consider linear realizations of the $\mathfrak{gl}(n)$ algebra and we use them to construct the corresponding star product, coproduct of momenta and twist as applications to a noncommutative space of $\mathfrak{gl}(n)$ type. Furthermore, we consider the dual realization and dual $\mathfrak{gl}(n)$ algebra, as well as a general realization of the $\mathfrak{gl}(n)$ algebra, coproduct of momenta and twists. These constructions were performed using methods proposed in \cite{MMSSt,remarks,BMMP,Ms} and in a Lie deformed phase space from twists in the Hopf algebroid approach in \cite{remarks,JMsalg,JKM,LMMPW,LMW}.

The plan of paper is as follows. In section 2 we introduce the generalized Heisenberg algebra and present linear realizations of the $\mathfrak{gl}(n)$ algebra. In section 3 the star product, and in section 4 the coproduct of momenta and twist are constructed. In section 5 the dual realization and dual $\mathfrak{gl}(n)$ algebra are considered. In section 6 a general realization of the $\mathfrak{gl}(n)$ algebra, the corresponding coproduct of momenta and two classes of twists are presented. These results can be applied to physical theories on noncommutative spaces of the $\mathfrak{gl}(n)$ type.

\section{Linear realizations of the $\mathfrak{gl}(n)$ algebra}

We introduce the generalized Heisenberg algebra generated with $n^2$ coordinates $x_{\mu\nu}$ and momenta $p_{\mu\nu}$, $\mu,\nu=1,...,n$, defined by
\begin{equation}
[x_{\mu\nu}, x_{\alpha\beta}]=0,\quad [p_{\mu\nu},p_{\alpha\beta}]=0, \quad [p_{\mu\nu},x_{\alpha\beta}]=-i\delta_{\mu\alpha}\delta_{\nu\beta}.
\end{equation}
Let us define the $\mathfrak{gl}(n)$ algebra generated with $\hat{x}_{\mu\nu}$, $\mu,\nu =1,...,n$
\begin{equation}
[\hat{x}_{\mu\nu},\hat{x}_{\lambda\rho}] = iu(\delta_{\mu\rho} \hat{x}_{\lambda\nu} -\delta_{\lambda\nu}\hat{x}_{\mu\rho})= iu C_{\mu\nu,\lambda\rho,\alpha\beta} \hat{x}_{\alpha\beta}, \quad u\in \mathbb{R},\label{gln1}
\end{equation}
where
\begin{equation}
C_{\mu\nu,\lambda\rho,\alpha\beta} = iu(\delta_{\mu\rho}\delta_{\lambda\alpha}\delta_{\nu\beta} -\delta_{\lambda\nu}\delta_{\mu\alpha}\delta_{\rho\beta}).
\end{equation}
are structure constants. Note that for $u=0$, $\hat{x}_{\mu\nu}$ commute among themselves.

One linear realization of $\hat{x}_{\mu\nu}$ is
\begin{eqnarray}
\hat{x}_{\mu\nu} &=& x_{\mu\nu}+ux_{\mu\alpha}p_{\nu\alpha},\label{1realiz}\\
\hat{x}_{\mu\nu} &=& x_{\alpha\beta}\varphi_{\alpha\beta,\mu\nu},
\end{eqnarray}
where
\begin{equation}
\varphi_{\alpha\beta,\mu\nu} = \delta_{\alpha\mu}\delta_{\beta\nu} +u\delta_{\alpha\mu}p_{\nu\beta}.
\end{equation}
Summation over repeated indices is understood throughout the whole paper.

Another linear realization of the $\mathfrak{gl}(n)$ algebra \eqref{gln1} is
\begin{eqnarray}
\hat{x}_{\mu\nu} &=& x_{\mu\nu} -ux_{\alpha\nu} p_{\alpha\mu},\\
\hat{x}_{\mu\nu} &=& x_{\alpha\beta} \varphi_{\alpha\beta,\mu\nu},
\end{eqnarray}
where
\begin{equation}
\varphi_{\alpha\beta,\mu\nu}=\delta_{\alpha\mu}\delta_{\beta\nu}- u\delta_{\beta\nu}p_{\alpha\mu}.
\end{equation}
This realization is related to the dual realization of the first realization \eqref{1realiz} with $u\mapsto -u$, see section 5.

\section{Star product}

Action $\bt$ is defined by
\begin{eqnarray}
x_{\mu\nu}\bt f(x) &=& x_{\mu\nu}f(x),\\
p_{\mu\nu} \bt f(x) &=& -i\frac{\partial f(x)}{\partial x_{\mu\nu}}.
\end{eqnarray}
Using $\bt$, it follows
\begin{eqnarray}
&& \hat{x}_{\mu\nu} \bt 1= x_{\mu\nu},\\
&& p_{\mu\nu} \bt \te ^{iqx}=q_{\mu\nu}\te ^{iqx},\quad \text{where}\; qx =q_{\alpha\beta}x_{\alpha\beta},\\
&& \te ^{ik_{\alpha\beta}\hat{x}_{\alpha\beta}} \bt \te ^{iq_{\gamma\delta}x_{\gamma\delta}} = \te ^{iJ_{\alpha\beta} (k,q,u)x_{\alpha\beta}},
\end{eqnarray}
for some $J_{\alpha\beta}(k,q,u)$.
If $k_{\alpha\beta}=0$, $J_{\mu\nu}(0,q,u)=q_{\mu\nu}$. If $q_{\mu\nu}=0$, $J_{\mu\nu}(k,0,u)=K_{\mu\nu}(k,u)$. If $u=0$, $J_{\mu\nu}(k,q,0)=k_{\mu\nu}+q_{\mu\nu}$.

From $J_{\mu\nu}(k,q,u)$, we can obtain the star product \cite{MMSSt,remarks}
\begin{equation}
\te ^{ikx}*\te ^{iqx}=\te ^{i\mathcal{D}(k,q,u)x},
\end{equation}
where
\begin{eqnarray}
&& \mathcal{D}_{\mu\nu}(k,q,u)=J_{\mu\nu}(K^{-1}(k),q,u),\\
&& K^{-1}_{\mu\nu}(K(k)) =K_{\mu\nu}(K^{-1}(k))=k_{\mu\nu}.
\end{eqnarray}
$J_{\mu\nu}(tk,q,u)$ is the unique solution of the partial differential equation \cite{MMSSt,BMMP,Ms}
\begin{equation}
\frac{\partial J_{\mu\nu} (tk,q,u)}{\partial t} =k_{\alpha \beta}\varphi_{\mu\nu,\alpha\beta} (J(tk,q,u)),
\end{equation}
with boundary condition $J_{\mu\nu}(0,q,0)=q_{\mu\nu}$. In our case,
\begin{equation}
\frac{\partial J_{\mu\nu}(tk,q,u)}{\partial t} =k_{\alpha\beta}(\delta_{\alpha\mu}\delta_{\beta\nu} +u \delta_{\alpha\mu} J_{\beta\nu}(tk,q,u))= k_{\mu\beta} (\delta_{\beta\nu} +uJ_{\beta\nu}(tk,q,u)).
\end{equation}
The differential equation in matrix form is
\begin{equation}
\frac{\partial J(tk,q,u)}{\partial t}=k (I+uJ(tk,q,u)).
\end{equation}
The solution in matrix form is
\begin{equation}
J(tk,q,u)= \frac{1}{u}(\te ^{utk}-I)+\te ^{utk}q,
\end{equation}
or, in terms of the components,
\begin{equation}
J_{\mu\nu}(tk,q,u) =\left(\te ^{utk}\right)_{\mu\alpha} q_{\alpha\nu} +\frac{1}{u}\left(-I+\te ^{utk}\right)_{\mu\nu}.
\end{equation}
For $t=0$
\begin{eqnarray}
&& J_{\mu\nu}(0,q,u)=q_{\mu\nu},\\
&& K_{\mu\nu}(k,u)=\frac{1}{u}\left(\te ^{uk}-I\right)_{\mu\nu},\\
&& K^{-1}_{\mu\nu}(k,u) =\frac{1}{u}(\ln(I+uk))_{\mu\nu}.
\end{eqnarray}
Hence
\begin{equation}
\mathcal{D}_{\mu\nu}(tk,q,u)=k_{\mu\nu}+q_{\mu\nu}+uk_{\mu\alpha}q_{\alpha\nu}.
\end{equation}
The star product is associative
\begin{equation}
(f*g)*h=f*(g*h).
\end{equation}

\section{Coproduct of momenta and twist}

The coproduct of momenta is
\begin{equation}\label{Dp}
\Delta p_{\mu\nu} =\mathcal{D}_{\mu\nu}(p\otimes 1,1\otimes p)= p_{\mu\nu}\otimes 1+1\otimes p_{\mu\nu}+ up_{\mu\alpha} \otimes p_{\alpha\nu}.
\end{equation}
This coproduct is coassociative $(\Delta \otimes 1)\Delta =(1\otimes \Delta)\Delta$.

In the Hopf algebroid approach \cite{JMsalg,JKM,LMMPW,LMW} we can write the twist \cite{remarks}
\begin{eqnarray}
\mathcal{F}^{-1} &=& :\,\exp(i(1\otimes x_{\alpha\beta})(\Delta -\Delta_0)p_{\alpha\beta})\,:\\
&=& \exp(-ip_{\alpha\beta}\otimes x_{\alpha\beta})\exp(iK^{-1}(p)_{\gamma\delta} \otimes \hat{x}_{\gamma\delta}),
\end{eqnarray}
where $:\,:$ denotes normal ordering, with the $x$s left from the $p$s. Using the expression~\eqref{Dp} for the coproduct $\Delta p$, we have
\begin{equation}
\mathcal{F}^{-1}_{1}=:\,\exp(iup_{\mu\nu}\otimes x_{\mu\alpha}p_{\nu\alpha})\,:
\end{equation}
and applying \cite{Ms}, Theorem 1 we get
\begin{equation}
\mathcal{F}^{-1}_1= \exp(i(\ln(I+up))_{\mu\nu}\otimes x_{\mu\alpha}p_{\nu\alpha}).
\end{equation}
Note that
\begin{equation}
K^{-1}_{\mu\nu}(p,u)=\frac{1}{u}(\ln(I+up))_{\mu\nu}.
\end{equation}
Here we introduce
\begin{equation}
Z_{\mu\nu}=(I+up)_{\mu\nu},
\end{equation}
and
\begin{equation}
L_{\mu\nu}=x_{\mu\alpha}p_{\nu\alpha}.
\end{equation}
$L_{\mu\nu}$ generate the $\mathfrak{gl}(n)$ algebra, with properties
\begin{eqnarray}
&& [Z_{\mu\nu},\hat{x}_{\lambda\rho}]=-iu\delta_{\mu\lambda}Z_{\rho\nu},\\
&& \Delta Z_{\mu\nu}=Z_{\mu\alpha} \otimes Z_{\alpha\nu},\\
&& \Delta(\ln Z)_{\mu\nu}=(\ln Z)_{\mu\nu} \otimes 1+1\otimes (\ln Z)_{\mu\nu},\\
&& [L_{\mu\nu},L_{\lambda\rho}]=i(\delta_{\mu\rho}L_{\lambda\nu}-\delta_{\lambda\nu}L_{\mu\rho}),\\
&& \Delta_0 L_{\mu\nu}=L_{\mu\nu}\otimes 1 +1\otimes L_{\mu\nu},\label{D0L}\\
&& \Delta_0 p_{\mu\nu}=p_{\mu\nu}\otimes 1+1\otimes p_{\mu\nu}.
\end{eqnarray}
Hence, the twist can be written as
\begin{equation}
\mathcal{F}^{-1}_1 =\exp(i(\ln Z)_{\mu\nu}\otimes L_{\mu\nu}).
\end{equation}
It is easy to check that
\begin{equation}
\hat{x}_{\mu\nu}=m\mathcal{F}^{-1}(\bt \otimes 1)(x_{\mu\nu}\otimes 1)= x_{\mu\nu} +ux_{\mu\alpha}p_{\nu\alpha},
\end{equation}
where $m$ is the multiplication map $m: A\otimes B\to AB$, and
\begin{equation}
\Delta p_{\mu\nu} =\mathcal{F} \Delta_0 p_{\mu\nu} \mathcal{F}^{-1}= \Delta_0 p_{\mu\nu} +up_{\mu\alpha} \otimes p_{\alpha\nu},
\end{equation}
where
\begin{equation}\nn
\Delta_0 p_{\mu\nu} =p_{\mu\nu} \otimes 1 +1 \otimes p_{\mu\nu}.
\end{equation}
Twist $\mathcal{F}_1$ satisfies the Drinfeld cocycle condition
\begin{equation}\label{cocycle}
(1\otimes \mathcal{F})(1\otimes \Delta_0)\mathcal{F}=(\mathcal{F} \otimes 1)(\Delta_0 \otimes 1)\mathcal{F}.
\end{equation}
The proof follows using the factorization property. Using $\Delta_0 L_{\mu\nu}$ \eqref{D0L}, we get
\begin{eqnarray}
&& (1\otimes \Delta_0) \mathcal{F}=\mathcal{F}_{12}\mathcal{F}_{13}=\mathcal{F}_{13}\mathcal{F}_{12},\\
&& (1\otimes \mathcal{F})(1\otimes \Delta_0)\mathcal{F} =\mathcal{F}_{23}\mathcal{F}_{12} \mathcal{F}_{13} =\mathcal{F}_{23} \mathcal{F}_{13} \mathcal{F}_{12}.
\end{eqnarray}
Using $\Delta(\ln Z)_{\mu\nu}=(\ln Z)_{\mu\nu}\otimes 1+1\otimes (\ln Z)_{\mu\nu}$, we get
\begin{equation}
(\mathcal{F}\Delta_0 \otimes 1)\mathcal{F} =\mathcal{F}_{13} \mathcal{F}_{23} \mathcal{F}_{12} =\mathcal{F}_{23}\mathcal{F}_{13}\mathcal{F}_{12}.
\end{equation}
Hence, the left and right side of \eqref{cocycle} become $\mathcal{F}_{23}\mathcal{F}_{13}\mathcal{F}_{12}$.

We note that the coproduct \eqref{Dp} may be related to a limiting case of the commutation relations for the braided Weyl algebra \cite{Gur1,Gur2} of a modified reflection equation algebra arising in the study of integrable systems.

\section{Dual realization and dual $\mathfrak{gl}(n)$ algebra}

Let us define the dual realization
\begin{eqnarray}
\hat{y}_{\mu\nu} &=& m\tilde{\mathcal{F}}^{-1}_1 (\bt \otimes 1)(x_{\mu\nu} \otimes 1) \\
&=& x_{\mu\nu} +ux_{\alpha\nu} p_{\alpha\mu}\\
&=& x_{\alpha\beta} \tilde{\varphi}_{\alpha\beta,\mu\nu},
\end{eqnarray}
where
\begin{equation}
\tilde{\varphi}_{\alpha\beta,\mu\nu} =\delta_{\alpha\mu}\delta_{\beta\nu} -u\delta_{\beta\nu} p_{\alpha\mu},
\end{equation}
and $\tilde{\mathcal{F}}$ denotes interchange of the left and right sides in the $\otimes$ product, $a\otimes b \to b\otimes a$. $\hat{y}_{\mu\nu}$ generate the dual $\mathfrak{gl}(n)$ algebra \eqref{gln1}, with $u\mapsto -u$.
\begin{eqnarray}
&& [\hat{y}_{\mu\nu}, \hat{y}_{\lambda\rho}] =-iu(\delta_{\mu\rho} \hat{y}_{\lambda\nu}- \delta_{\lambda\nu}\hat{y}_{\mu\rho}),\label{glnd}\\
&& [Z_{\mu\nu},\hat{y}_{\lambda\rho}]=-iu\delta_{\rho\nu}Z_{\mu\lambda},
\end{eqnarray}
and
\begin{equation}
[\hat{x}_{\mu\nu},\hat{y}_{\lambda\rho}]=0.
\end{equation}
The star product induced with the dual realization $\hat{y}_{\mu\nu}$ is constructed as in section 3. The partial differential equation for $\tilde{J}_{\mu\nu} (tk,q,u)$ is
\begin{eqnarray}
\frac{\partial \tilde{J}_{\mu\nu}(tk,q,u)}{\partial t} &=& k_{\alpha\beta} \tilde{\varphi}_{\mu\nu,\alpha\beta} (\tilde{J}(tk,q,u))\\
&=& k_{\mu\nu} +u\tilde{J}_{\mu\alpha} (tk,q,u)k_{\alpha\nu}.
\end{eqnarray}
The differential equation in matrix form is
\begin{equation}
\frac{\partial \tilde{J}(tk,q,u)}{\partial t}=(I+u\tilde{J}(tk,q,u))k.
\end{equation}
The solution in matrix form is
\begin{equation}
\tilde{J}(tk,q,u)=\left(\frac{\te ^{utk}-I}{u}\right) +q\te ^{utk},
\end{equation}
and
\begin{equation}
\tilde{J}_{\mu\nu}(tk,q,u)=q_{\mu\alpha}\left(\te ^{utk}\right)_{\alpha\nu} +\left(\frac{\te ^{utk}-I}{u}\right)_{\mu\nu}.
\end{equation}
For $t=0$,
\begin{eqnarray}
&& \tilde{J}_{\mu\nu}(0,q,u)=q_{\mu\nu},\\
&& \tilde{K}_{\mu\nu}(k,u) =\frac{1}{u}\left(\te ^{uk}-I\right)_{\mu\nu} =K_{\mu\nu}(k,u),\\
&& \tilde{K}^{-1}_{\mu\nu}(k,u) =\frac{1}{u}(\ln (I+uk))_{\mu\nu}=K^{-1}_{\mu\nu}(k,u).
\end{eqnarray}
Hence,
\begin{equation}
\tilde{\mathcal{D}}_{\mu\nu}(k,q,u)=k_{\mu\nu} +q_{\mu\nu}+uk_{\alpha\nu}q_{\mu\alpha}.
\end{equation}
The star product is associative.

The coproduct of $p_{\mu\nu}$ is
\begin{equation}\label{tDp}
\tilde{\Delta}p_{\mu\nu}=p_{\mu\nu}\otimes 1 +1\otimes p_{\mu\nu} +up_{\alpha\nu}\otimes p_{\mu\alpha},
\end{equation}
where $\tilde{\Delta}$ denotes the interchange of the left and right side in the tensor product with respect to $\Delta$, $a \otimes b\to b\otimes a$. The coproduct is coassociative.

The corresponding twist $\mathcal{F}_2$ in the Hopf algebroid approach \cite{remarks} (sec.4) and the coproduct $\tilde{\Delta}p_{\mu\nu}$ \eqref{tDp} lead to
\begin{eqnarray}
\mathcal{F}^{-1}_2 &=& :\,\exp\left(iup_{\nu\alpha} \otimes x_{\mu\alpha}p_{\mu\nu}\right)\,:\\
&=& \exp\left(i(\ln(I+up))_{\mu\nu} \otimes x_{\alpha\nu} p_{\alpha\mu}\right).
\end{eqnarray}
Note that
\begin{equation}
\tilde{K}^{-1}_{\mu\nu}(p,u)=\frac{1}{u}\left(\ln(I+up)\right)_{\mu\nu} =\frac{1}{u}\left(\ln Z\right)_{\mu\nu}.
\end{equation}
We define
\begin{equation}
\tilde{L}_{\mu\nu}=x_{\alpha\nu} p_{\alpha\mu}.
\end{equation}
$\tilde{L}_{\mu\nu}$ generate the $\mathfrak{gl}(n)$ algebra
\begin{eqnarray}
&& [\tilde{L}_{\mu\nu},\tilde{L}_{\lambda\rho}]=-i(\delta_{\mu\rho} \tilde{L}_{\lambda\nu}- \delta_{\lambda\nu}\tilde{L}_{\mu\rho}),\\
&& \Delta_0 \tilde{L}_{\mu\nu}=\tilde{L}_{\mu\nu}\otimes 1 +1\otimes \tilde{L}_{\mu\nu}.
\end{eqnarray}
Hence, the twist can be written as
\begin{equation}
\mathcal{F}^{-1}_2=\exp \left(i(\ln Z)_{\mu\nu}\otimes \tilde{L}_{\mu\nu}\right).
\end{equation}
It is easy to check that
\begin{eqnarray}
\hat{y}_{\mu\nu}&=& m\mathcal{F}^{-1}_2 (\bt \otimes 1)(x_{\mu\nu}\otimes 1)\\
&=& x_{\mu\nu} +ux_{\alpha\nu} p_{\alpha\mu},
\end{eqnarray}
and
\begin{eqnarray}
\tilde{\Delta}p_{\mu\nu} &=& \mathcal{F}_2 \Delta_0 p_{\mu\nu} \mathcal{F}^{-1}_2\\
&=& \Delta_0 p_{\mu\nu} +up_{\alpha\nu} \otimes p_{\mu\alpha}.
\end{eqnarray}
Twist $\mathcal{F}_2$ satisfies the Drinfeld cocycle condition. From the duality property, it follows
\begin{eqnarray}
\hat{x}_{\mu\nu} &=& m\tilde{\mathcal{F}}_2^{-1}(\bt \otimes 1)(x_{\mu\nu} \otimes 1)\\
&=& x_{\mu\nu} +ux_{\mu\alpha} p_{\nu\alpha},
\end{eqnarray}
and\begin{equation}
\Delta p_{\mu\nu} =\tilde{\mathcal{F}}_2 \Delta_0 p_{\mu\nu} \tilde{\mathcal{F}}^{-1}_2.
\end{equation}
It is important to note that twists $\mathcal{F}_1$ and $\tilde{\mathcal{F}}_2$ are different, but lead to the same $\hat{x}_{\mu\nu}$ and the same coproduct $\Delta$. Analogously, twists $\tilde{\mathcal{F}}_1$ and $\mathcal{F}_2$ are different, but they lead to the same $\hat{y}_{\mu\nu}$ and coproduct $\tilde{\Delta}$. Also, $\hat{y}_{\mu\nu}(-u)$ gives another linear realization of the $\mathfrak{gl}(n)$ algebra \eqref{gln1}, different from $\hat{x}_{\mu\nu}$ \eqref{1realiz}. The dual realization with respect to $\hat{y}_{\mu\nu}(-u)$ is $\hat{x}_{\mu\nu}(-u)$, generating $\mathfrak{gl}(n)$, with $u\mapsto-u$.

\section{General realization of the $\mathfrak{gl}(n)$ algebra, coproduct of momenta and two classes of twists}

A general realization of the generators $\hat{x}_{\mu\nu}$ and $\hat{y}_{\mu\nu}$ generating the $\mathfrak{gl}(n)$ algebra \eqref{gln1} and the $\mathfrak{gl}(n)$ algebra \eqref{glnd}, respectively, can be obtained using a similarity transformation on the coordinates $x_{\mu\nu}$ and momenta $p_{\mu\nu}$. Let us define new coordinates and momenta as
\begin{eqnarray}
x'_{\mu\nu} &=& \te ^{x_{\alpha\beta}S_{\alpha\beta}(p,u)+T(p,u)} x_{\mu\nu} \te ^{-(x_{\alpha\beta}S_{\alpha\beta}(p,u)+T(p,u))},\\
p'_{\mu\nu} &=& \te ^{x_{\alpha\beta}S_{\alpha\beta}(p,u)+T(p,u)} p_{\mu\nu} \te ^{-(x_{\alpha\beta}S_{\alpha\beta}(p,u)+T(p,u))}=\Lambda_{\mu\nu}(p,u).
\end{eqnarray}
Hence, using the inverse relations, we obtain
\begin{eqnarray}
\hat{x}_{\mu\nu} &=& x'_{\alpha\beta}\varphi'_{\alpha\beta,\mu\nu}(p',u) +\chi'_{\mu\nu}(p',u),\label{hxpreko'}\\
\hat{y}_{\mu\nu} &=& x'_{\alpha\beta} \tilde{\varphi}'_{\alpha\beta,\mu\nu} (p',u) +\tilde{\chi}'_{\mu\nu}(p',u),\label{hypreko'}
\end{eqnarray}
where $\varphi'_{\alpha\beta,\mu\nu}$, $\tilde{\varphi}'_{\alpha\beta,\mu\nu}$, $\chi'_{\mu\nu}$ and $\tilde{\chi}'_{\mu\nu}$ depend on $S(p,u)$ and $T(p,u)$. Note that, after inserting the realizations of $\hat{x}_{\mu\nu}$ \eqref{hxpreko'} and $\hat{y}_{\mu\nu}$ \eqref{hypreko'}, relations $[\hat{x}_{\mu\nu},\hat{y}_{\lambda\rho}]=0$ are satisfied. Furthermore $\Delta p'_{\mu\nu}$ is obtained from
\begin{equation}
\Delta p'_{\mu\nu} = \Delta (\Lambda_{\mu\nu}(p,u))= \Lambda_{\mu\nu}(\Delta p,u)\vert_{p =\Lambda^{-1}(p',u)},
\end{equation}
i.e. expressing $p$ as $\Lambda^{-1}(p',u)$. A similar statement holds for $\tilde{\Delta}p'_{\mu\nu}$.

From $\Delta p'_{\mu\nu}$ and $\tilde{\Delta}p'_{\mu\nu}$ we easily obtain $\mathcal{D}'_{\mu\nu}(k,q,u)$ and $\tilde{\mathcal{D}}'_{\mu\nu}(k,q,u)$ that define the corresponding star products. Namely, 
\begin{eqnarray}
&& \Delta p'_{\mu\nu} =\mathcal{D}'_{\mu\nu} (p'\otimes 1,1\otimes p'),\label{Dp'}\\
&& \tilde{\Delta}p'_{\mu\nu} =\tilde{\mathcal{D}}'_{\mu\nu}(p'\otimes 1,1\otimes p'),\\
&& \te ^{ikx'}*_1\te ^{iqx'} =\te ^{i\mathcal{D}'(k,q,u)x'}\\
&& \te ^{ikx'}*_2 \te ^{iqx'} =\te ^{\tilde{\mathcal{D}}'(k,q,u)x'},\\
&& \Delta p'_{\mu\nu} (\bt \otimes \bt)(\te ^{ikx'}\otimes \te ^{iqx'}) =\mathcal{D}'_{\mu\nu} (k,q,u)\left( \te ^{ikx'}\otimes  \te ^{iqx'}\right),\label{Dp'btbt}\\
&& \tilde{\Delta}p'_{\mu\nu} (\bt \otimes \bt)(\te ^{ikx'} \otimes \te ^{iqx'}) =\tilde{\mathcal{D}}'_{\mu\nu} (k,q,u)\left( \te ^{ikx'}\otimes \te ^{iqx'}\right),\\
&& \Delta_0 p'_{\mu\nu}=p'_{\mu\nu} \otimes 1+1\otimes p'_{\mu\nu}.
\end{eqnarray}
Generally, $f*_1g=g*_2f$, for arbitrary $f(x')$ and $g(x')$. A consistency check is the relation between $\hat{x}_{\mu\nu}$ and $\Delta p'_{\alpha\beta}$
\begin{equation}
\hat{x}_{\mu\nu}= x'_{\mu\nu} +ix'_{\alpha\beta} m(\Delta -\Delta_0)p'_{\alpha\beta} (\bt \otimes 1) (x'_{\mu\nu}\otimes 1) +\chi'_{\mu\nu}(p',u),
\end{equation}
and similarly for $\hat{y}_{\mu\nu}$.

The family of twists $\mathcal{F}_{1}(s)$ in the Hopf algebroid approach corresponding to the general realization of $\hat{x}_{\mu\nu}$ \eqref{hxpreko'}, for $\chi'_{\mu\nu}(p',u)=0$, $T(p,u)=0$ is
\begin{equation}
\mathcal{F}^{-1}_1(s) =:\,\exp\left( i((1-s)(1\otimes x'_{\alpha\beta}) +sx'_{\alpha\beta} \otimes 1)(\Delta -\Delta_0)p'_{\alpha\beta}\right) \,:,
\end{equation}
where $\Delta p'_{\mu\nu}=p'_{\mu\nu}\otimes 1+1 \otimes p'_{\mu\nu}$, and $s\in \mathbb{R}$. We point out that although these twists are different for different $s$, they define the same star product for all $s$. Namely,
\begin{eqnarray}
\te ^{ikx'}*_1 \te ^{iqx'} &=& m\mathcal{F}^{-1}_1 (s) (\bt \otimes \bt) (\te ^{ikx'} \otimes \te ^{iqx'})\nn \\
&=& m\exp \left(i((1-s)1\otimes x'_{\alpha\beta}+ s1\otimes x'_{\alpha\beta})(\Delta -\Delta_0) p'_{\alpha\beta})(\mathcal{D}'_{\alpha\beta}(k,q,u) -k_{\alpha\beta} -q_{\alpha\beta})\right)\left(\te ^{ikx'}\otimes \te ^{iqx'}\right)\nn \\
&=& \te ^{ix_{\alpha\beta}\mathcal{D}'_{\alpha\beta}(k,q,u)}
\end{eqnarray}
does not depend on $s$ due to \eqref{Dp'btbt}, the definition of normal ordering and the multiplication map $m$. Since $\Delta p'_{\mu\nu}=\mathcal{D}'(p'\otimes 1,1\otimes p')$, \eqref{Dp'}, the coproduct $\Delta p'_{\mu\nu}=\mathcal{F}_1(s) \Delta_0 p'_{\mu\nu} \mathcal{F}^{-1}_1(s)$ does not depend on the parameter $s$. Similarly, $\hat{x}_{\mu\nu}= m\mathcal{F}^{-1}_1(s) (\bt \otimes 1)(x'_{\mu\nu}\otimes 1)$ does not depend on $s$. Analogously, we can define the family of twists $\mathcal{F}_2(s)$ corresponding to the general realization of $\hat{y}_{\mu\nu}$ \eqref{hypreko'} with $\chi'_{\mu\nu}(p',u)=0$ and $T(p,u)=0$
\begin{equation}
\mathcal{F}^{-1}_2(s) =:\, \exp\left( i(1-s)(1\otimes x'_{\alpha\beta} +sx'_{\alpha\beta} \otimes 1)(\Delta-\Delta_0)p'_{\alpha\beta}\right)\,:.
\end{equation}
Note that $\mathcal{F}_1(s)=\tilde{\mathcal{F}}_2(1-s)$.

\section*{Aknowledgements}

This work has been started during the Franco-Croatian joint program "COGITO", the project N 24829NH "Syst\`emes Int\'egrables et Structures Non-commutatives" supported by EGIDE. We thank D. Gurevich for inspiring discussions. Z. \v{S}. also acknowledges hospitality of LAREMA (Angers) and LAMAV (Valenciennes).

\end{document}